# MOBILE CLOUD COMPUTING IN HEALTHCARE USING DYNAMIC CLOUDLETS FOR ENERGY-AWARE CONSUMPTION

Manoj Muniswamaiah[1] and Dr. Charles Tappert[2]

[1]Seidenberg School of CSIS, Pace University, White Plains, New York
{mm42526w, ctappert}@pace.edu

*ABSTRACT*

*Mobile cloud computing (MCC) has increasingly been adopted in healthcare industry by healthcare professionals (HCPs) which has resulted in the growth of medical software applications for these platforms. There are different applications which help HCPs with many important tasks. Mobile cloud computing has helped HCPs in better decision making and improved patient care. MCC enables users to acquire the benefit of cloud computing services to meet the healthcare demands. However, the restrictions posed by network bandwidth and mobile device capacity has brought challenges with respect to energy consumption and latency delays. In this paper we propose dynamic energy consumption mobile cloud computing model (DEMCCM) which addresses the energy consumption issue by healthcare mobile devices using dynamic cloudlets.*

## 1. INTRODUCTION

Healthcare professionals (HCPs) use applications which are deployed on Mobile cloud computing (MCC) platform for better decision making and patient care. MCC enables users to take the advantage of cloud computing services. It is the combination of mobile computing, networking and cloud computing. Even though MCC has been adopted widely it faces the issue of energy been wasted by mobile devices when it cannot establish connection with wireless network and it keeps searching for it. In this paper we use dynamic energy consumption mobile cloud computing model (DEMCCM) which addresses energy consumption issue by healthcare mobile devices by using dynamic cloudlets which optimizes the usage of cloud infrastructure services. DEMCCM uses dynamic programming model to enable cloudlets within changing environments to assist MCC. The applications running on these mobile platforms rely on the speed and connectivity of the wireless communication. Also, instability in the wireless network shortens the battery life of these mobile devices.

DEMCCM offers a unique technique to avoid energy wastage by HCPs mobile devices when the networking environment is unstable. This model basically focuses on the communication between HCPs devices and cloud servers. Figure [1] represents the conceptual model of DEMCCM. The cloudlets provide an operating platform for dynamic search execution. HCPs mobile devices send requests to the nearest cloudlets before they have been redirected to the cloud servers.

The cloudlets choose the cloud server for better service performance based on the dynamic programming. The cloud server chosen by the dynamic model cloudlets is the optimal solution provided by the model which is expected to avoid energy wastage of the HPCs mobile devices. The main contribution of this paper is the dynamic energy consumption implementation for cloudlets to improve the performance and save energy. Our model is an extension of work done by (Keke Gai, Meikang Qiu, Hui Zhao, LixinTao, Ziliang Zong et al.,2015) [1].

## 2. TERMINOLOGIES

*Green Technology*
Green Technology is the usage of computers and its resources to be echo friendly. The model implemented in this paper enables green computing by reducing the energy consumption. It focuses on cloudlet techniques such as virtualization and dynamic programming. By applying DEMCCM we aim to reduce the energy consumption on mass HCPs devices without weakening the cloud service performance.

*Mobile Cloud Computing*
Mobile Cloud Computing integrates cloud computing with mobile devices to make them resourceful in terms of memory, storage and computation power. Although cloud is useful for computation, traditional offloading techniques cannot be used for HCPs mobile devices directly because these techniques are generally energy-unaware and face wireless network bandwidth issues [2]. The below table shows the issues between cloud and mobile cloud computing services.

| Issues | Cloud Computing | Mobile Cloud Computing |
| --- | --- | --- |
| Device Energy | No | Yes |
| Security | Yes | Yes |
| Bandwidth | No | Yes |
| Location Awareness | No | Yes |
| Context Awareness | No | Yes |
| Mobility | No | Yes |
| Network Connectivity | No | Yes |
| Bandwidth Utilization Cost | No | Yes |

Table 1: Issues in Cloud and Mobile Cloud Computing

*Computation Offloading Decision Factors*
When a local execution on a mobile device consumes more resource, computation offloading can serve as a major performance booster by transferring resource intensive computational tasks to an external platform like a cluster, or a grid, or a cloud. The decision of computational offloading is an intricate process and the decision itself is influenced by different factors. The entities that influence the computation offloading decisions are HCPs preference, connection, smart phone and cloud service.

A *HCP* may prefer to enable computation offloading. If the HCP's objective is to protect their patient's data and are not certain about the integrity of the offloaded data, then they can disable computation offloading.

*Network Connection* can affect the decision of computation offloading. Modern mobile devices used by HCPs can communicate through different networking interfaces such as Wi-Fi, 3G / 4G and each of these interfaces can have their own limitations. Wi-Fi technology, for instance, can provide higher bandwidth and shorter delays. Cellular connections like 3G / 4G provide lower bandwidth and suffer from higher delays. It consumes more amount of energy for data transmission. If both of these connections are available to a user, then they can prefer to use Wi-Fi connection. However, HCPs needs to be mobile themselves and Wi-Fi connection is not feasible in all the places. Thus forcing them to switch to cellular network connection 3G / 4G, for which they are charged based on the bandwidth usage. Hence, from connection point of view, the decision to enable or disable computation offloading can be influenced by network bandwidth, delay and cost.

*Smartphone* plays a significant role in computation offloading decision. They have achieved great development in terms of hardware resources. Today's smart phones are equipped with high performance processors, memory, sensors and storage. For instance, the Apple Iphone X features a 14.73 cm display and runs on iOS v11.0.1 operating system. HCPs who use such powerful smartphones are less likely to opt for mobile cloud support as compared to HCPs who have low performance smartphones that runs out of resources quickly.

*Application Model* should also be considered while opting for computation offloading. Each mobile cloud application is different in terms of both their design and objectives. Some of the applications may be designed with an objective to reduce the energy consumption or increase application performance for devices that do not have enough resources locally. The application models may be different in terms of context awareness, application partitioning, code availability on the cloud, profilers and overhead. Application nature can also influence computation offloading decision. Applications like GPS, sensors or clinical communication apps require local hardware resources. In order to execute these applications on the cloud, the application needs to be partitioned into off loadable, that can be retained, and non-off loadable, that needs to be moved to cloud components. Similarly, applications whose input data size is larger can be difficult to be moved to cloud, since transferring large input data incurs higher turnaround time and consume higher energy for communication.

*Cloud Service* is important in computation offloading. If the clinical mobile cloud application requires computation offloading, then the cloud service must have runtime support and be rich in resources in order to gain the advantage of computation offloading [3].

*Cloudlets*
Cloudlets form an intermediate layer in between HCPs devices and cloud servers. A cloudlet is a self-management systems which is used for strengthening communication between HPCs devices and cloud servers by reducing latency. The cloudlets are deployed with dynamic programming and the requests from application running on the HCPs devices will reach the nearest cloudlet. Before the requests are directed to the cloud servers, calculations regarding the resource allocations are in the dynamic programming algorithm of the cloudlet. It allows to determine the physical machine available for serving the request and also predict whether the users should switch to other cloudlets. The main purpose of the model is to save energy consumption and reduce latency. Once the request is received by cloudlet, business logic is applied and executed using dynamic programming [4].

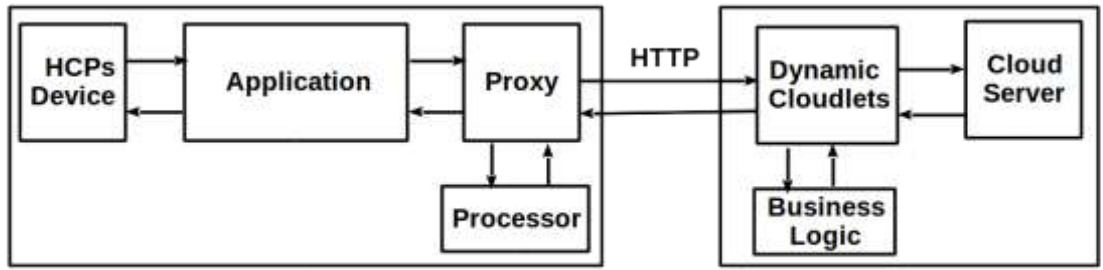

Fig 1: Fundamental Concepts of DEMCCM

## 3. HEALTHCARE PROFESSIONALS DEVICES AND APPLICATION SCENARIO

Healthcare professionals use applications running on mobile devices for decision making and patient care. In a traditional way these devices communicate with cloud servers directly. This results in the devices and cloud servers to be connected until signal becomes weak or gets disconnected. DEMCCM dynamically selects the nearby cloud server for processing of the request and returns the response to the application saving energy consumption.

HCPs devices access the nearest Cloudlet A that is assumed to be the nearest one at the time of serving the request. Later if Cloudlet A determines that Cloudlet B can offer better performance based upon the location of the device and networking condition the connection will be switched to Cloudlet B which helps the users to get better performance from the Cloudlet B. Cloudlets are responsible for searching the cloud server to serve the request. The communication between cloudlets layer and cloud servers are responsible for serving requests and responses dynamically as shown in Figure 2.

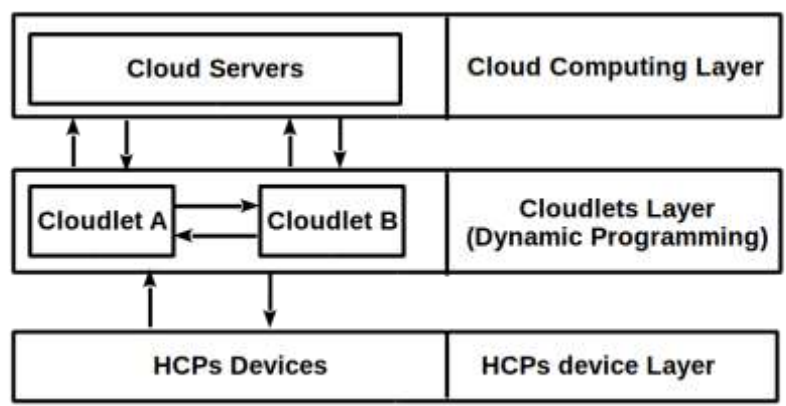

Fig 2: Conceptual flow of DEMCCM Model

## 4. DYNAMIC PROGRAMMING ALGORITHM

1. DEMCCM *algorithm*

Definition: GECM: *Green Energy Consumption Minimization problem during communication between HCPs devices and cloud server by adopting cloudlets*. Given set of cloudlets c, $\forall c \in N \; \forall c > 1$, where N is a natural number.

The total cost, $C_{TotalCost}$(T,c), consists of two components, the first component provides the energy consumption of HCPs devices, cloudlets and cloud servers which occurs on one of the request route defined as $En_i$(t,c). Also second component provides the performance level on one of the route c for a specific time unit t defined as $Pf_i$(t,c). This is basically to make sure that the quality of service and performance provided while choosing the energy aware route is optimal. The goal of the model is to produce minimum energy consumption under specific time.

Minimize $C_{TotalCost}$(T,c)   $(\forall T \in R \; ; \; \forall c \in N)$    [1]

The total energy consumption is a sum of the products of energy of each of the model and the performance levels.

$C_{TotalCost}(T,c) = \sum_{x=1}^{n} f(Pf_i(t,c), En_i(t,c))$   $(\forall i \in N; \; \forall c \in N; \; \forall t \in R^+)$   [1]

Using the dynamic programming DEMCCM model aims to select the most efficient route between HCPs devices and the cloud severs. Dynamic programming algorithm generates recursion which drills down in to smaller sub-set of problems to find the solution to the whole problem. It selects best solution for each sub-problem during each iteration [5]. We use dynamic programming to choose the cloudlet that can provide better service. As shown in the below formulation we can obtain cloudlet that has minimum energy consumption. energy cost at each cloudlet is given by $Pf_i$(t,c), $En_i$(t,c) and the total cost is formulated as f($Pf_i$(t,c), $En_i$(t,c))

*EnergyCloudlet($C_{i,t}(M_1)$) = CostperCL[$C_{i,t}(M_1)$]*
*EnergyCloudlet($C_{i,t}(M_2)$) = CostperCL[$C_{i,t}(M_2)$]*
  .
*EnergyCloudlet($C_{i,t}(M_n)$) = CostperCL[$C_{i,t}(M_n)$]*    [1]

Dynamic programming algorithm for cloudlets

**Input:** $Pf_i(t,c), En_i(t,c), T_i(c), M_i(c)$, and $Node(c)$
**Output:** Green energy consumption for a specified time

```
for c ← 1 to Node(c) do
  for m ← 1 to M_i(c) do
    for t_τ ← to T_i(c) do
      Pf_m(t_τ, c) ← ∑_{t=1}^{t_τ} Pf_m(t,c)
      En_m(t_τ, c) ← En_m(t_τ, c)
      Remove the worst pair after comparison
    end for
  end for
end for
for t_τ ← 1 to T_i(c) do
end for
return all CostperCL[c_{i,t}(M_x)]
```

| Notations | Definitions |
|---|---|
| c | Represents cloudlet node |
| r | Method route to be used, $\forall r \in N$ |
| $M_i$ | Method route is the cloudlet route to be used, $\forall i \in N$ |
| t | Defines latency or timing cost for each node |
| T | Unit time |
| $Pf_i(t,c)$ | Performance at the time t and cloudlet c |
| $En_i(t,c)$ | Energy cost at the time t and cloudlet c |
| © | Represents performance and energy cost $(Pf, En) = (Pf_i, En_i) © (Pf_j, En_j); Pf = Pf_i \times Pf_j; En = En_i + En_j$ |
| $C_{i,t}(M_i(c))$ | Attributes of one cloudlet with two variables, i and t, and the corresponding $M_i(c)$ |
| CostperCL | Performance and energy consumption of each cloudlet |
| $C_{TotalCost}(T,c)$ | Total energy consumption considering service performance within a specific schedule period |
| $f(Pf_i(t,c), En_i(t,c))$ | Total cost function with two variables $Pf_i(t,c), En_i(t,c)$ |

Table 2: Elements used for formulation of algorithm

## 5. EVALUATION

Evaluation is based on simulation which is a mathematical deduction between DEMCCM model and the traditional cloud computing method. The workload has been examined by DEMCCM-sim which was developed in the lab. It takes the input data and generates result using dynamic programming. The performance is measured using network latency time. Experiments were conducted to determine whether DEMCCM is applicable for operating interval associated with specific timing constraints and also to evaluate whether there is optimized solution for each timing constraint when two or more routes were deployed. The second experiment was conducted to test the operability of DEMCCM in a complicated environment. Table 3 shows the result of energy consumption and performance of cloudlets and Table 4 shows performance results under various time constraints using DEMCCM model.

| T | (P,E) | (P,E) | (P,E) | (P,E) | (P,E) |
|---|---|---|---|---|---|
| 1 | (0.163,44.0) | | | | |
| 2 | (0.226,54.0) | (0.219,54.0) | (0.122,41.0) | | |
| 3 | (0.226,50.0) | (0.134,39.0) | (0.232,54.0) | (0.114,44.0) | (0.210,59.0) |
| 4 | (0.237,39.0) | (0.335,44.0) | (0.317,58.0) | (0.19,44.0) | (0.229,49.0) |
| 5 | (0.357,50.0) | (0.348,44.0) | (0.229,46.0) | (0.445,56.0) | (0.356,49.0) |
| 6 | (0.360,49.0) | (0.437,58.0) | (0.445,44.0) | (0.267,36.0) | (0.349,42.0) |
| 7 | (0.59,51.0) | (0.469,42.0) | (0.478,49.0) | (0.378,78.0 ) | (0.429,47.0) |
| 8 | (0.176,33.0) | (0.416,48.0) | (0.57,41.0) | (0.526,46.0) | (0.283,33.0) |
| ⋮ | | | | | |
| 23 | (0.781,23.0) | (1.03,31.0) | (0.5,24.0) | | |
| 24 | (0.89,22.0) | (1.9,21.0) | (1.2,37.0) | | |
| 25 | (0.83,27.0) | (1.1,35.0) | (1.2,28.0) | | |

Table 3 : Performance and energy consumption between different cloudlets

| Node | M1 | | | M2 | | |
| --- | --- | --- | --- | --- | --- | --- |
| | T1 | P1 | E1 | T2 | P2 | E2 |
| 0 | 1 | 0.7 | 7 | 2 | 0.6 | 6 |
| | 2 | 0.1 | 7 | 3 | 0.2 | 6 |
| 1 | 1 | 0.8 | 6 | 2 | 0.7 | 3 |
| | 2 | 0.3 | 6 | 3 | 0.3 | 3 |
| 2 | 1 | 0.8 | 9 | 2 | 0.8 | 4 |
| | 3 | 0.2 | 9 | 4 | 0.2 | 4 |
| 3 | 2 | 0.7 | 8 | 5 | 0.8 | 5 |
| | 3 | 0.3 | 8 | 7 | 0.3 | 5 |
| 4 | 2 | 0.6 | 7 | 4 | 0.8 | 6 |
| | 5 | 0.4 | 7 | 5 | 0.2 | 6 |
| 5 | 1 | 0.7 | 8 | 2 | 0.7 | 2 |
| | 3 | 0.3 | 8 | 4 | 0.3 | 2 |
| 7 | 1 | 0.8 | 6 | 3 | 0.8 | 3 |
| | 4 | 0.4 | 6 | 5 | 0.2 | 3 |

Table 4: Minimum cost computed under various time constraints for DEMCCM model

## 6. CONCLUSION

This paper proposed DEMCCM model to reduce the energy consumption by mobile devices. The performance was also evaluated against different time constraints. We would like to apply this model to other industries as part of future research.

## REFERENCES

[1]  http://webpage.pace.edu/kg71231w/docs/jnca1.pdf
[2]  Kumar, Karthik, and Yung-Hsiang Lu. "Cloud computing for mobile users: Can offloading computation save energy?." *Computer* 4 (2010): 51-56.
[3]  Othman, Mazliza, Sajjad Ahmad Madani, and Samee Ullah Khan. "A survey of mobile cloud computing application models." *IEEE Communications Surveys & Tutorials* 16.1 (2013): 393-413.
[4]  Sanaei, Zohreh, et al. "Heterogeneity in mobile cloud computing: taxonomy and open challenges." *IEEE Communications Surveys & Tutorials* 16.1 (2013): 369-392.
[5]   Hong, Kirak, et al. "Mobile fog: A programming model for large-scale applications on the internet of things." *Proceedings of the second ACM SIGCOMM workshop on Mobile cloud computing*. ACM, 2013.